# Analysing Magnetism Using Scanning SQUID Microscopy

P. Reith[1], X. Renshaw Wang (王骁)[1,2,*], H. Hilgenkamp[1]

[1]MESA+ Institute for Nanotechnology, University of Twente, P.O. Box 217, 7500 AE, Enschede, The Netherlands

[2]School of Physical and Mathematical Sciences & School of Electrical and Electronic Engineering, Nanyang Technological University, 50 Nanyang Avenue, 639798 Singapore

*Email: renshaw@ntu.edu.sg

**ABSTRACT**

**Scanning superconducting quantum interference device microscopy (SSM) is a scanning probe technique that images local magnetic flux, which allows for mapping of magnetic fields with high field and spatial accuracy. Many studies involving SSM have been published in the last decades, using SSM to make qualitative statements about magnetism. However, quantitative analysis using SSM has received less attention. In this work, we discuss several aspects of interpreting SSM images and methods to improve quantitative analysis. First, we analyse the spatial resolution and how it depends on several factors. Second, we discuss the analysis of SSM scans and the information obtained from the SSM data. Using simulations, we show how signals evolve as a function of changing scan height, SQUID loop size, magnetization strength and orientation. We also investigated 2-dimensional autocorrelation analysis to extract information about the size, shape and symmetry of magnetic features. Finally, we provide an outlook on possible future applications and improvements.**

## Introduction

Scanning superconducting quantum interference device microscopy (scanning SQUID microscopy, or SSM) is part of the scanning probe microscopy (SPM) family[1]. Its main purpose is to image the local magnetic flux on sample surfaces. In general, SQUIDs are the most sensitive magnetometers currently available[2,3], making SSM an excellent instrument to image magnetism with high spatial as well as field accuracy.

Several other methods to image magnetism have been developed over the years, some of which we will mention here. The most common one is magnetic force microscopy (MFM), which measures the force between a magnetized scan probe and the sample surface. Scanning Hall probe microscopy[4] (SHPM) makes use of the appearance of a Hall voltage in a magnetic field. The magnitude of the Hall voltage provides a measure for the strength of the magnetic field. There is also a variety of optics-based methods with the ability to measure magnetic properties[5-9].

Beside the unprecedented magnetic field accuracy, SSM techniques have an added strength in that they can be applied to a wide variety of magnetic phenomena. Typical objects of interest are superconducting vortices[1, 10-12] and ferromagnetic surfaces[13-17], but SSM can also be used for imaging the magnetic field originating from current distributions[1, 18-20] and local susceptibility using an auxiliary field loop[21-23]. SSM has also been applied to measuring fractional flux quanta[24-26] and the coexistence of superconductivity and ferromagnetism[27]. Local manipulation of the surface magnetism has also been demonstrated[28]. Finally, SSM can also be a tool to image magnetic contaminations on a non-magnetic surface, or a surface with a known magnetic field structure (e.g., current leads or predefined magnetic structures). One disadvantage of SSM is the spatial resolution. Compared to MFM, which can achieve a resolution down to a few tens of nanometers[29, 30], SSM has only recently entered the sub-micrometre regime[31, 32].

Up to now, SSM analysis of these kinds of features was mostly done in a qualitative manner. With increasing use of SSM, quantitative analysis of SSM data is in demand. This will lead to a better understanding of the data and underlying physics. Previously, limitations in spatial resolution made the fitting of data to models and extracting sample characteristics difficult. Here, we will discuss ways to improve the quantitative analysis of SSM images, and show how certain image features depend on material or scan parameters. The first section of this work will discuss the spatial resolution of SSM, how it depends on different factors and we will show that it is not as straightforward as in other SPM techniques. The second part will focus on SSM imaging of magnetic features. The discussion will focus mostly on magnetic dipoles and ferromagnetic surfaces.

The discussions in this publication are written in the context of SSM, but are relevant to other techniques as well. The considerations on spatial resolution can be translated to other imaging techniques. Considering, for example, the Hall structure used in SHPM to image magnetism, this Hall structure with a finite size will influence the system's resolution, similar to the effect from a pickup loop in SSM. Looking outside magnetic imaging, an X-ray beam used in scanning nano-X-ray diffraction[33] (SNXRD) will also have a finite diameter. The discussion on analysis of data in this publication (or at the least, the mathematics underlying them) has broader applications as well. For example, the discussion on autocorrelation can also be used for SNXRD to give information about a sample's physical structure.

SI-units and formulae for magnetism are used in this paper.

## Working Principle of Scanning SQUID Microscopy

A direct-current (dc) SQUID as used in SSM systems consists of a superconducting ring containing two Josephson junctions biased with a constant current. From basic quantum mechanics, one can find that the total magnetic flux $\Phi$ threading a superconducting ring must be quantized in units of the flux quantum $\Phi_0 = 2.0678 * 10^{-15}$ Tm$^2$:

$$\Phi = n \times \frac{h}{2e} = n\Phi_0, \tag{1}$$

where $h$ is Planck's constant, $e$ the elementary charge and $n$ an integer. By adding the two Josephson junctions, Equation (1) changes to the following:

$$\frac{1}{2\pi}(\varphi_1 - \varphi_2) + \frac{\Phi}{\Phi_0} = n, \tag{2}$$

where $\varphi_1$ and $\varphi_2$ are the phase drops across each Josephson junction, and $n$ an integer. From this, we can see that in a dc SQUID, the flux $\Phi$ is related to the phase. Combining this with the Josephson equation

$$I = I_c \sin(\varphi), \tag{3}$$

we see that flux couples to the critical current.

In practice, dc SQUIDs in SSM systems are often operated in the voltage state[34]. There, the flux $\Phi$ relates to the voltage across the dc SQUID in a periodic, sine-like manner. This is the preferred modus operandi since voltage is measured more easily than the critical current. To combat the non-linearity of the flux-voltage relation, a feedback loop is applied. The dc SQUID is put at a certain working point on the flux-voltage relation (usually where $d\Phi/dV$ is largest), and the feedback circuit will keep the system at that working point when moved by an external magnetic flux. The current passed through the feedback coil causes a voltage drop across an accompanying feedback resistance, which is then a measure for the flux threading the dc SQUID.

The dc SQUID is typically extended with a pickup loop[1, 35] (Figure 1a), which has a well-defined area for the flux to penetrate while the rest of the SQUID is magnetically shielded. This is to reduce the influence of the external magnetic field on the SQUID itself. A SQUID is only sensitive to the magnetic field component perpendicular to the SQUID plane (note that this does not mean it cannot image objects with magnetic moment parallel to the SQUID plane, as we will discuss later). If we assume the sensor to be parallel to the surface, this component would be the $z$-component.

Most of the analysis in this work will be based on the magnetic dipole equation, given by

$$\vec{B} = \frac{\mu_0}{4\pi} \left[ \frac{3\vec{r}(\vec{m} \cdot \vec{r})}{r^5} - \frac{\vec{m}}{r^3} \right], \qquad (4)$$

where $\mu_0$ is the magnetic permeability of free space, $\vec{r} = \{x; y; h\}$ the distance vector from source to observer with magnitude $r$, and $\vec{m} = \{m_x; m_y; m_z\}$ the magnetic moment of the dipole with magnitude $m$. From this, we can determine the $z$-component of a dipole field:

$$B_z = \frac{\mu_0}{4\pi} \left[ \frac{3h(m_x x + m_y y + m_z h)}{r^5} - \frac{m_z}{r^3} \right]. \qquad (5)$$

SSM data is typically displayed as either magnetic flux or magnetic field. Since an SSM setup measures magnetic flux, displaying data as such will be more accurate, because it involves fewer conversions. However, displaying data as magnetic field is more practical, because it can be more easily related to other properties of the sample. The conversion between field and flux involves the area of the pickup loop, which we will discuss in the next section.

## Spatial Resolution

We define the in-plane area of the object of interest to be the $xy$-plane, and the direction normal to the sample surface as the $z$-direction (Figure 1a). The sensor is held at a sample-sensor distance $h$, with an angle $\theta$ between it and the $xy$-plane. We then divide the scanned area into pixels, where one pixel corresponds to one data point. A pixel has a set area $A_p = \Delta x \Delta y$, where $\Delta x$ and $\Delta y$ are the step size in the $x$ and $y$ directions respectively (Figure 1b).

Because of the high precision of linear motors or piezo actuators, the step size is usually reduced to values below the dimensions of the pickup loop (Figure 1b). Therefore, the flux $\Phi_s$ that threads the pickup loop is not equal to the flux $\Phi_p$ that threads one pixel. If the magnetic field $B$ is assumed to be constant across the surface of the pickup loop $A_s$, we can state:

$$B = \frac{\Phi_p}{A_p} = \frac{\Phi_s}{A_s} \rightarrow \Phi_p = \frac{A_p}{A_s} \Phi_s. \qquad (6)$$

We can see that to obtain the corrected value for the flux through a pixel, a correction factor $A_p/A_s$ has to be multiplied with the measured flux.

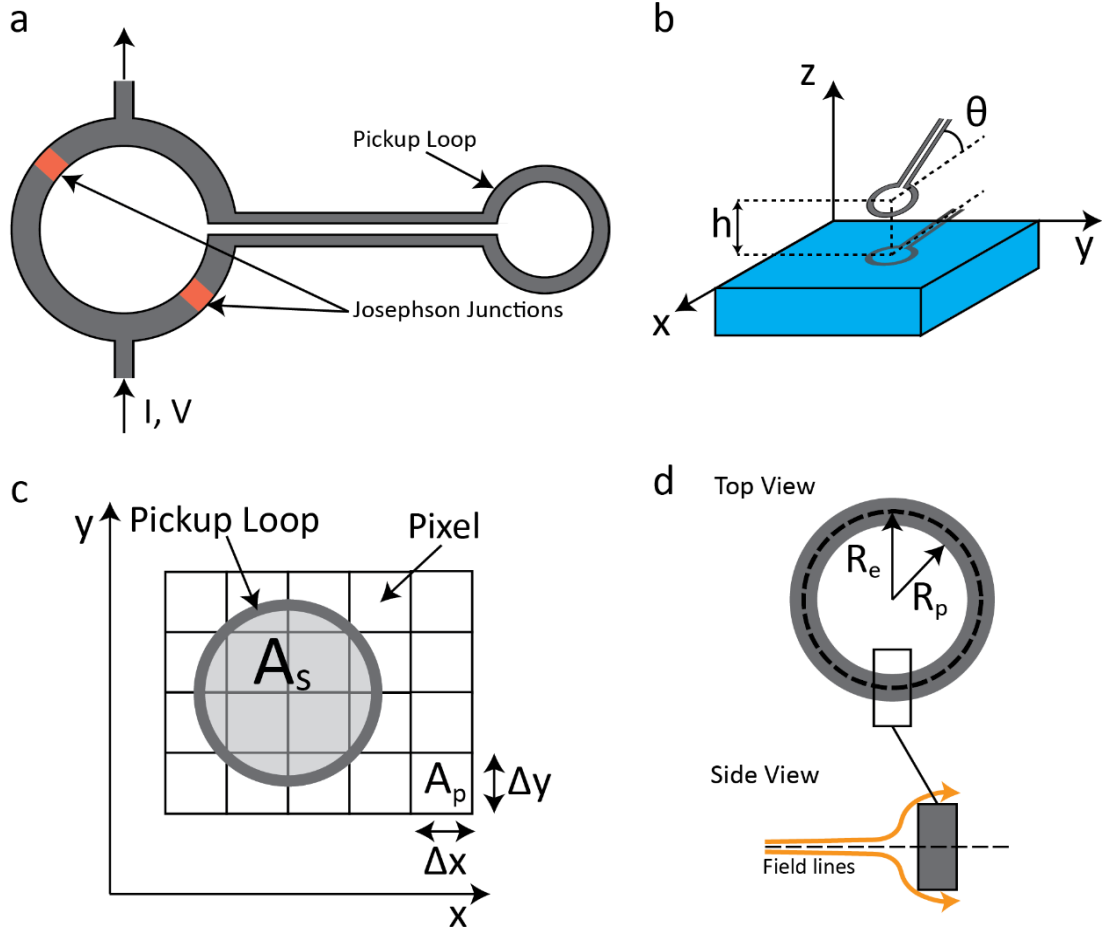

Figure 1. (a) Schematic depiction of a SQUID, extended with a pickup loop. (b) Definition of the SSM coordinate system and the scanning height $h$ and scanning angle $\theta$. (b) Comparison between the SQUID pickup loop area $A_s$ and the pixel area $A_p$. (c) Schematic depiction of the behavior of magnetic field lines near a superconductor and the influence on the effective diameter $R_e$ of the pickup loop.

Since the SQUID circuitry (including the pickup loop) is made out of superconducting material, the flux threading the pickup loop is subject to flux focusing. Because superconductors are perfect diamagnets (barring Abrikosov vortices), magnetic field lines that would normally penetrate the material of the pickup loop are instead bent around the material through the inner area of the pickup loop. This causes an effective enhancement of the pickup loop area, leading to an increased response of the system. This means that when converting flux to magnetic field, the effective area of the pickup loop is slightly larger than the area enclosed by the pickup loop. A rigorous analysis was done by Ketchen and Kirtley [36]. As a first approximation, one can assume that incoming flux lines will be split halfway to bend around a superconducting material (see Figure 1c). If one takes a superconducting ring, then the effective diameter $R_e$ of that ring will be the physical diameter $R_p$ plus half the width of the ring. $A_s$ is therefore increased to an effective area $A_e$, and Equation (6) becomes

$$\Phi_p = \frac{A_p}{A_e}\Phi_s. \tag{7}$$

$A_e$ can be obtained experimentally by measuring a single Abrikosov vortex[20, 37]. Since the total flux coming from a vortex must equal $1\Phi_0$, the correction to the physical area can be obtained by summing all the data points and dividing by the flux quantum. This assumes all flux coming from a vortex is imaged by the SSM.

Next, we will discuss the spatial resolution of the SSM system. One of the key features of SSM setups, and SQUIDs in general, is their high precision in measuring magnitude of magnetic flux, and discussions on field sensitivity can be found elsewhere[3, 38, 39]. However, as mentioned, their spatial resolution is lacking compared to, for example, MFM.

Spatial resolution is the ability of a system to discern between different features. In most publications, the spatial resolution is not mentioned, only the geometry of the sensor and the sample-sensor distance is given[11, 13, 19, 21-24, 27, 32, 40-46]. Knowing the SSM spatial resolution is key in understanding the data obtained from an experiment. As we will show below, the spatial resolution of an SSM setup is a complex combination of both the pickup loop diameter and the sample-sensor distance.

A magnetic field will have a certain magnetic field profile $B(x, y)$ (Figure 2a). When imaged by SSM, the resulting flux profile $\Phi$ will be broadened due to the finite size $R_e$ of the pickup loop (Figure 2b). The amount of broadening depends linearly on $R_e$.

Conversely, as the sample-sensor distance increases, the field profile will broaden (Figure 2a), which in turn causes $\Phi$ to broaden as well. This is independent of the size of the pickup loop. As we can see, both factors influence the final flux profile $\Phi$. This means that increasing the height, the pickup loop size, or both, will cause broadening of $\Phi$ until two neighbouring features can no longer be distinguished (Figure 2c). Therefore, defining the spatial resolution based only on one of these two factors gives an incomplete picture of a system's capabilities.

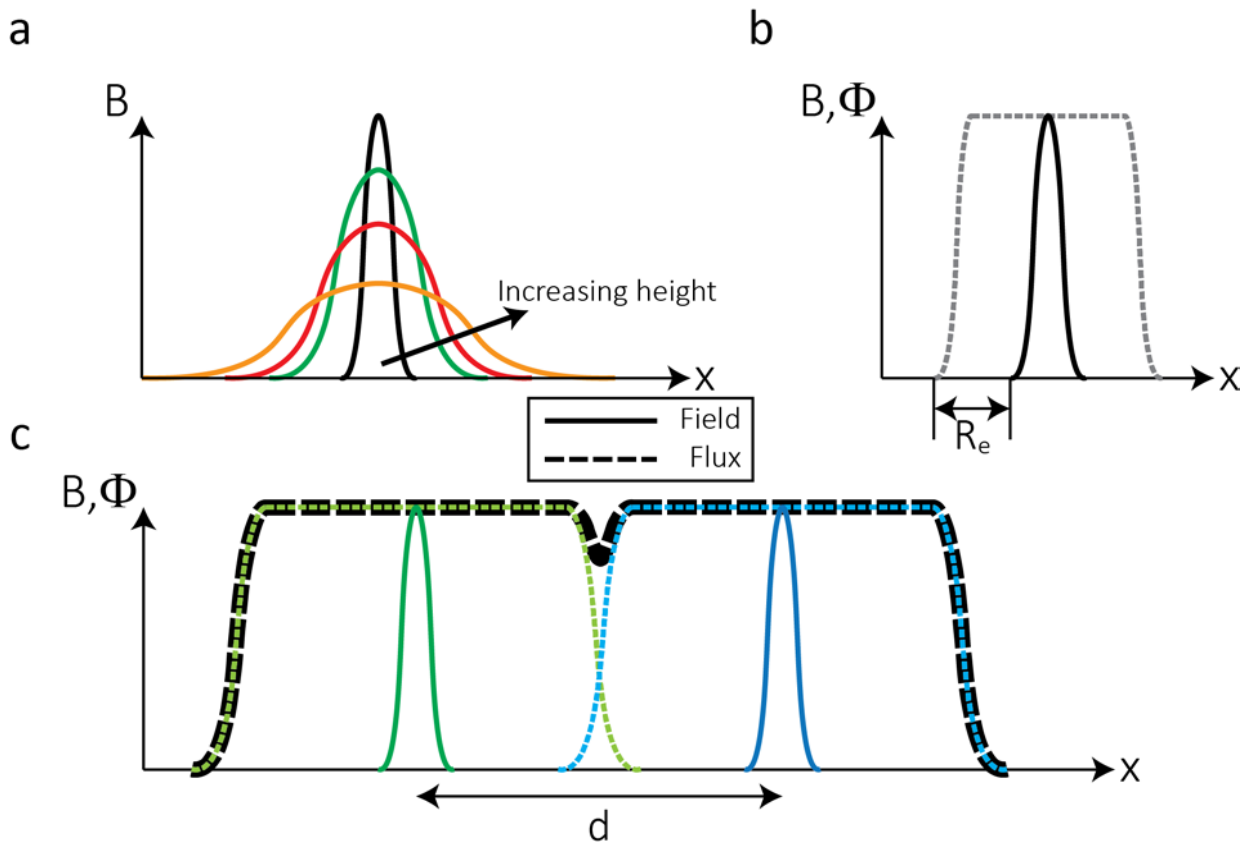

**Figure 2.** (a) Magnetic profile $B$ for some magnetic feature as a function of height. (b) Flux profile $\Phi$ (dashed) obtained by measuring delta function $B$ (solid) with a pickup loop of radius $R_e$. (c) Two magnetic features, represented as delta functions (blue and green solid curves) separated by a distance $d$ with overlapping flux profiles (light blue and light green dashed curves, respectively) produce a combined flux profile (black dashed curve) that can obscure the individual features.

To properly determine the spatial resolution of an SSM system, Kirtley et al. suggested a Rayleigh-like criterion[35]. Their definition states that two Abrikosov vortices are resolved if the intensity of the flux profile between them drops to 81% of the maximum value, with the spatial resolution being equal to the distance between these two features. To illustrate, Kirtley et al. obtained a spatial resolution of 11.2 μm with their 10 μm pickup loop[35].

The problem with basing a definition of the spatial resolution on an Abrikosov vortex is that current SSM techniques are approaching resolutions that are on the scale of the physical size of the vortex. This means that, when determining the resolution, the vortex size will have to be taken into account. This can be problematic, since the physical size of a vortex is not trivial and depends on several different factors such as film thickness[47].

We therefore suggest a new definition of spatial resolution, based on an in-plane point-dipole (Figure 3a). Such a dipole will produce an out-of-plane field that will have a two extrema (Figure 3b). The separation $s$ of these extrema depends purely on the height $h$ at which it is measured, and the diameter $d$ of the sensor. That is, in the limit that both go to zero, $s$ goes to zero. We now define the spatial resolution to be equal to $s$.

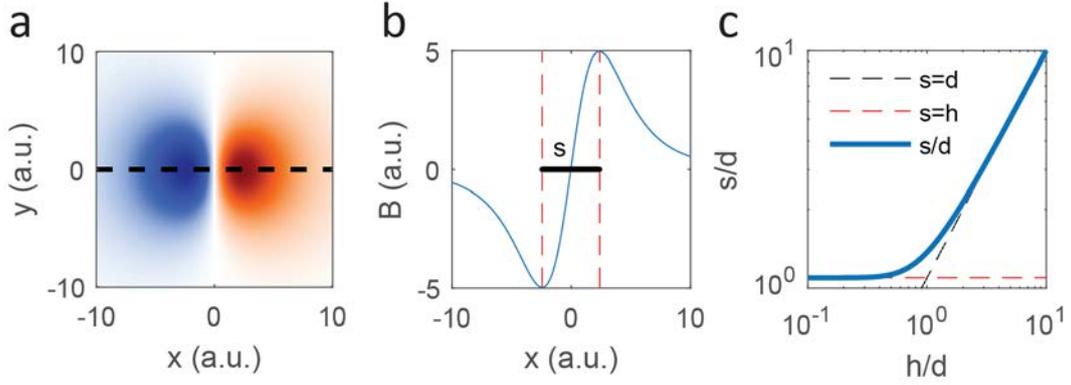

Figure 3. Definition of spatial resolution based on a point-dipole. (a) Out-of-plane field image of a point-dipole at the origin with a magnetic moment aligned along the x-axis. (b) Magnetic field profile indicated in (a). The red dashed lines indicate the two extrema, with the separation distance $s$ indicated with the black solid line. (c) The normalized resolution $s/d$ as a function of $h/d$ (blue solid line), with the limits $s = d$ (red dashed) and $s = h$ (black dashed).

To determine $s$, we can derive the equation to determine the location of the two extrema as a function of scan height $h$ and pickup loop radius $R = d/2$. This equation can easily be derived from Equation (5) and leads to

$$\frac{\frac{s}{2}+R}{\left(\left(\frac{s}{2}+R\right)^2+h^2\right)^{5/2}} = \frac{\frac{s}{2}-R}{\left(\left(\frac{s}{2}-R\right)^2+h^2\right)^{5/2}} \qquad (8)$$

Figure 3c shows the normalized resolution $s/d$ as a function of the normalized height $h/d$. Also indicated are the two limiting cases: $s = d$ for low values of $h/d$, and $s = h$ for high values of $h/d$. Within 1% error, one can take $s = d$ for $h/d < 0.37$, and $s = h$ for $h/d > 5.5$. Equation (8) and Figure 3c allow for determining the spatial resolution on any scale.

We have chosen this definition because an in-plane dipole field is easily recognizable and relatable for users of SSM systems or readers of SSM-related literature.

# Imaging of Magnetic Features

In this section we will discuss the analysis of data provided by an SSM system. Again, we would like to emphasize that, although this section is written from the perspective of SSM, it can be applied to other imaging techniques, including techniques outside of magnetic imaging.

The field of magnetic features smaller than the spatial resolution can often be simplified to a point-dipole field shown in Equation (4). Simulated fields of point-dipoles as they would be imaged by an SSM setup are shown in Figure 4 for different inclination angles $\theta$. As shown, the typical double-lobe picture of a dipole field is only visible under certain angles. The data can be fitted to the dipole equation to obtain the magnetic moment $\vec{m}$ of the magnetic feature.

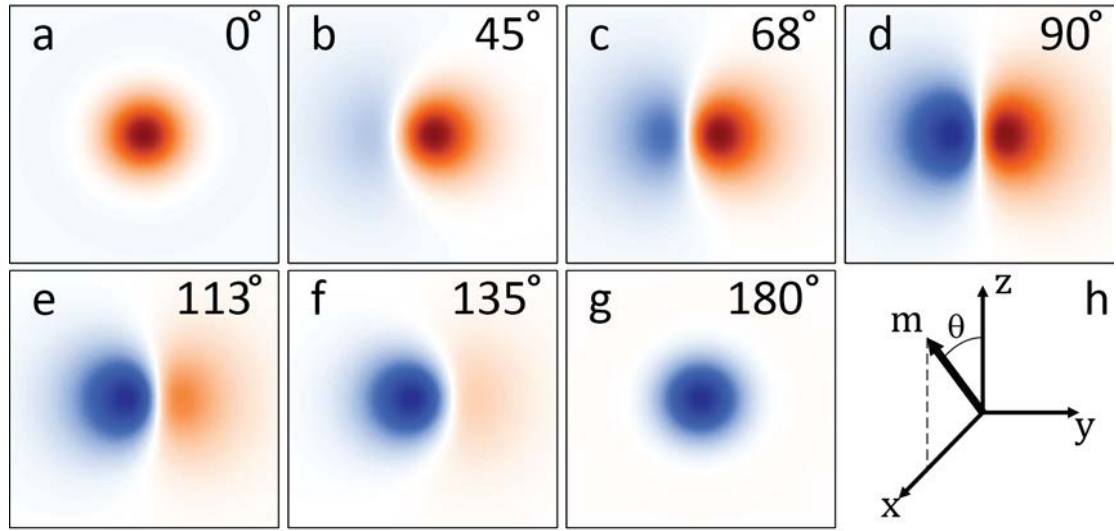

**Figure 4.** (a-g) Simulated magnetic field intensity images of dipoles with changing inclination angle $\theta$ with the $z$-axis. (h) Coordinate system of (a-g).

Figure 4d also shows how SSM images in-plane magnetic moment. In order to close the field lines, the magnetic field must rotate through the out-of-plane direction to reverse and close the loop. This is why, in the in-plane case of Figure 4d, the magnetic field signal is strongest a small distance away from the point-dipole.

To go to ferromagnetism from an individual dipole, we will have a short look at a collection of dipoles. Figure 5 shows what several dipoles located close together would look like when imaged by an SSM. This could be the case, for example, when imaging a sample containing magnetized particles. We can see that for a few dipoles (Figure 5a-c), the individual dipoles can still be seen clearly enough. But as the number of dipoles increases (Figure 5d-e), the resulting image quickly becomes too complicated to accurately determine the number of dipoles. One can use fitting algorithms to determine the properties of the dipoles using Equation (5), but since every dipole has 6 degrees of freedom (3 for position and 3 for magnetic moment), this will use a lot of computing power even for a small number of dipoles, and the solution is not necessarily unique.

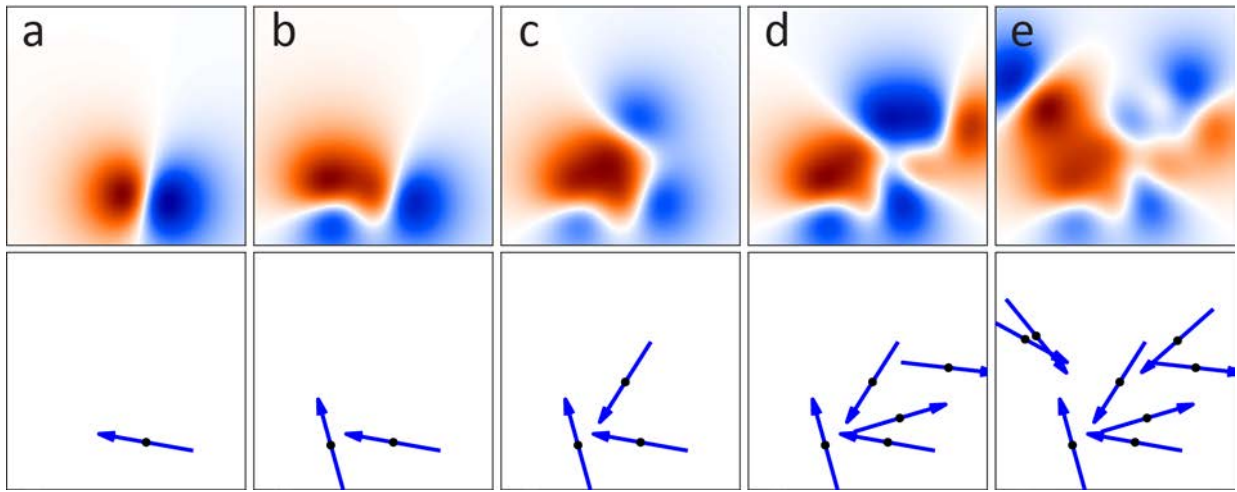

Figure 5. (a-e) Simulations for 1, 2, 3, 5 and 8 point dipoles (black dots). The arrows indicate the direction of the dipole magnetic moment.

Next, we will discuss ferromagnetic surfaces. SSM has been applied more and more to image surface ferromagnetism[13-17, 27], following the rise of thin film science. Apart from global magnetic field and moment data, properties of interest are usually magnetic domain size and shape, as well as any preferential magnetic moment orientation.

To complement this discussion, simulations have been done to highlight certain aspects of ferromagnetism. The simulations shown in this section have been carried out as follows: The simulated region is divided into domains, each with a certain magnetic moment and orientation. The domains are subdivided into pixels, which are treated as point dipoles. The fields of these dipoles are summed to get the fields of the domains, which are finally combined to get the simulated SSM image. The simulation parameters are based on earlier research performed in our group[16] and are indicative of typical SSM images (see Figure 6 for a comparison).

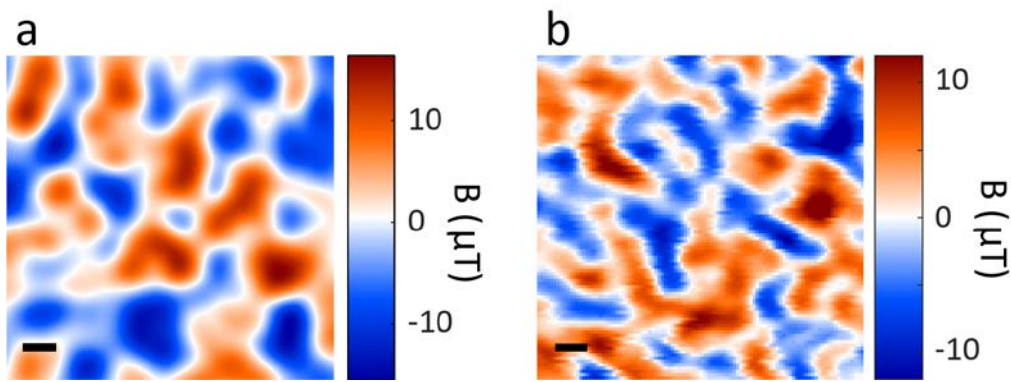

Figure 6 - Comparison between a simulated ferromagnetic surface and an SSM scan of a ferromagnetic surface. (a) Simulated surface using known parameters for the thin film ferromagnet $LaMnO_3$ and our SSM setup. (b) SSM data for a thin film of $LaMnO_3$ on Niobium-doped $SrTiO_3$. Scalebars indicate 10 $\mu$m.

Because ferromagnetic materials typically have a domain structure, this will be reflected in the SSM image (see Figure 6). However, it is important to note that the domain structure visible on the SSM image may not be representative of the underlying domains in the material itself. As the spatial resolution worsens due to increasing height, the field of different domains will be summed to form a weaker, averaged field. This is shown in Figure 7a-d, where a domain structure is simulated at different heights. One can clearly see that the domain structure as seen by the SSM changes: domains become larger and fewer in number due to averaging, with lower overall field values. The inhomogeneity in magnetic field will persist because of the random nature of domains. This means one has to be careful about making comments on the domain size as seen in SSM images.

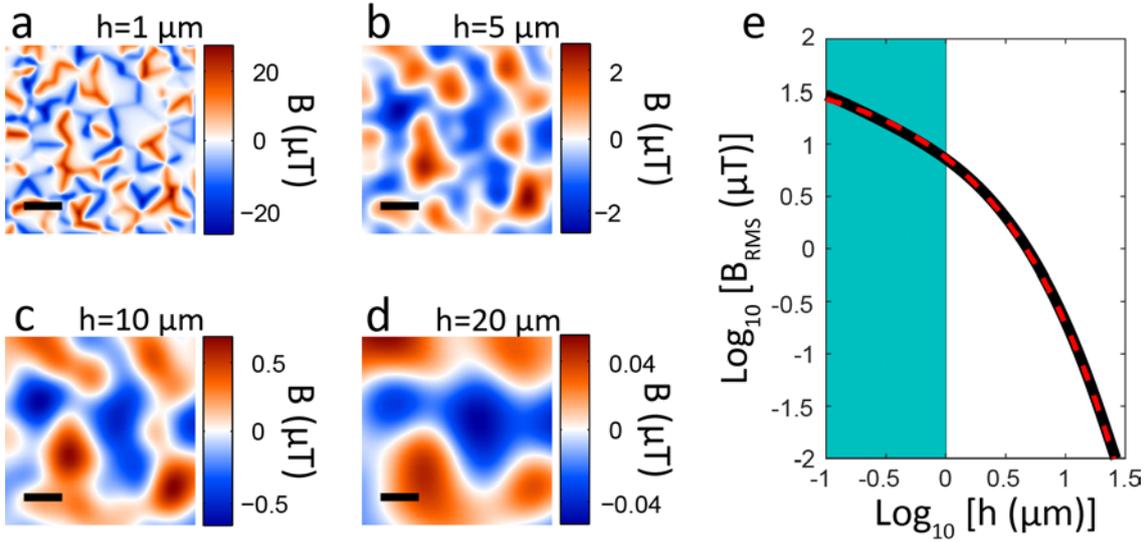

Figure 7. Simulated ferromagnetic surface containing an in-plane magnetized domain structure at different height: (a) 1μm; (b) 5 μm; (c) 10 μm; (d) 20 μm. The scale bar indicates 10 μm. (e) Dependence of $B_{RMS}$ on the scanning height (black solid line). The shaded area shows a region of relatively low decrease in signal. The red dashed line is an empirical fit to the simulation results of the form $\log_{10} B_{RMS} = -413h^{0.45} - 4.27$.

We can also see that the signal is strongest at the edges of the domain. As discussed before, this is due to the field lines having to reverse direction in order to close. This causes the out-of-plane component of the magnetic field to be the largest at domain boundaries (where it reverses), and weakest in the center of domains (where the field is almost completely in-plane).

To quantify the overall strength of the magnetic field on a ferromagnetic surface, we propose to use the root-mean-squared (RMS) field value $B_{RMS}$ as a figure of merit:

$$B_{RMS} = \sqrt{\frac{1}{n}\sum_{i=1}^{n}(B_i - \bar{B})^2}, \qquad (10)$$

where $n$ is the number of data points, $B_i$ is the field value at point $i$, and $\bar{B}$ is the average magnetic field. This allows for comparison between samples when certain sample preparation or scanning parameters are changed.

Figure 7e shows the dependence of $B_{RMS}$ on the scanning height $h$. We can see that that $B_{RMS}$ decreases faster with increasing height. The shaded region in Figure 7e shows the $h$-values for which the decrease in $B_{RMS}$ is relatively low, indicating an optimal scanning height. We see that above 1 μm, the value of $B_{RMS}$ rapidly decreases, indicating a large loss of information. An empirical fit was found of the form

$$\log_{10} B_{RMS} = -413h^{0.45} - 4.27. \qquad (11)$$

The above function serves only to give an approximation to the obtained curve and has no theoretical basis.

Another aspect to investigate is how $B_{RMS}$ changes with magnetic moment direction and strength. Figures 8a and c highlight the stark difference between in-plane and out-of-plane magnetic moment. This is caused by the out-of-plane field component being largest near the domain walls in the case of in-plane magnetic moment, and largest near the center of a domain for out-of-plane magnetic moment.

As the height increases, the visual difference that is present at low height (compare Figure 8a to Figure 8c), becomes negligible (Figures 8b and d). This complicates making claims about the orientation of the magnetic moments if the spatial resolution is not sufficient.

Figure 8e shows the dependence of $B_{RMS}$ on the magnetic moment for the in-plane and out-of-plane oriented cases. Since the domain structure in a simulation is randomly generated, multiple simulations were averaged to get the results shown. We can see that $B_{RMS}$ depends linearly on the magnetic moment $m$, which corresponds to the linear relation between $B$ and $m$ for a point dipole as shown in Equation (5).

Looking at the out-of-plane series, we see a larger coefficient between $B_{RMS}$ and the magnetic moment compared to the in-plane series. The larger values are a natural result of the field lines being mostly aligned along the $z$-axis, which is the component that is picked up by the SQUID.

Another factor that influences this coefficient is the ratio between the surface area and the boundary of a domain. As discussed before, in-plane magnetism will be visible at the edges of a domain, meaning more edges will lead to higher $B_{RMS}$ values. Conversely, for out-of-plane magnetism, the field is visible above the domain, and zero at the domain edge, meaning fewer boundaries cause higher $B_{RMS}$ values.

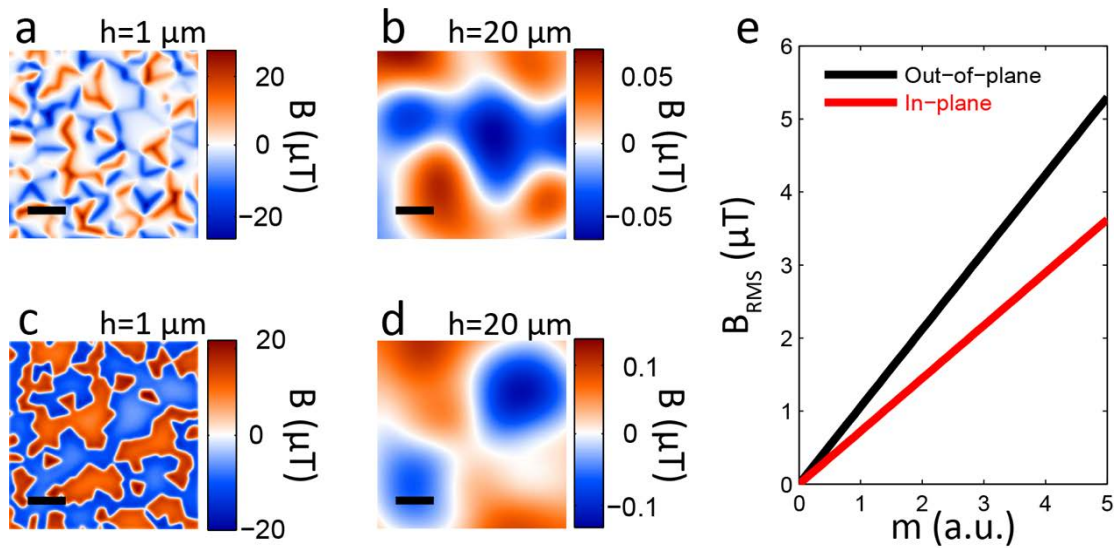

**Figure 8.** Simulated 100x100 µm² ferromagnetic surfaces with in-plane (a,b) and out-of-plane (c,d) orientation, at scan height of 1 µm (a,c) and 20 µm (b,d). (e) Relation between $B_{RMS}$ and the magnetic moment $m$.

Finally, we can look at the effect of imaging using a pickup loop with a non-zero size. As discussed before, this will cause a broadening of the measured signal due to averaging across the whole loop. Figure 9a shows the same simulation shown in Figure 7a and Figure 8a. Figure 9b shows this same simulation, but then calculated with a pickup loop with a diameter of 10 µm. This is done by, for each point, taking the average of all data points within a diameter $d$ of that point.

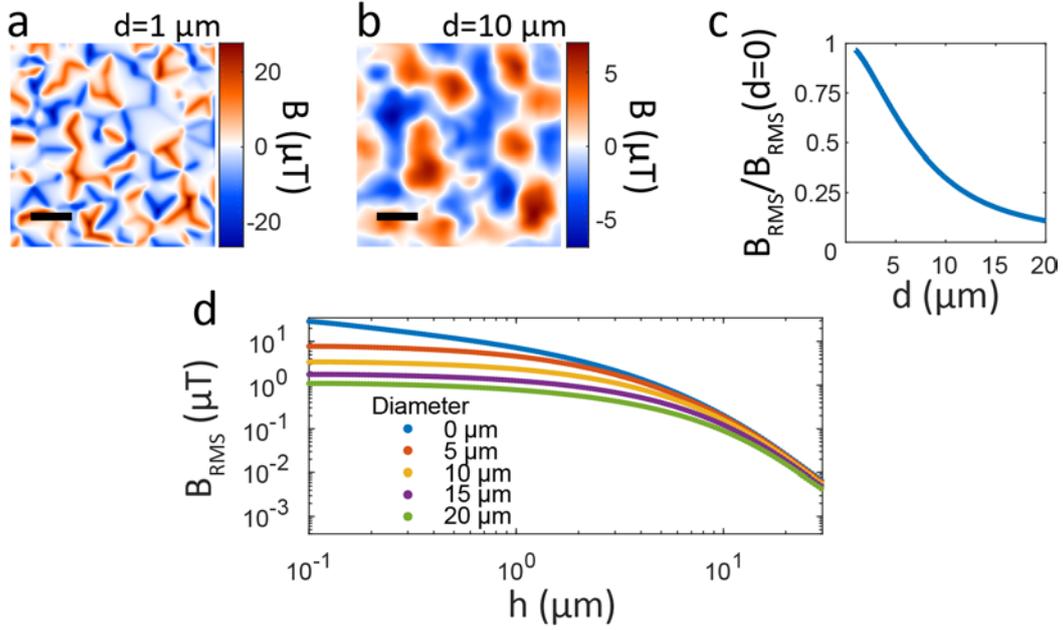

Figure 9. Influence of pickup loop diameter on imaging ferromagnetism. (a) Simulated ferromagnetic domain structure as imaged with a pickup loop of size zero. (b) Same as in (a), but with a pickup loop with a diameter of 10 μm. (c) $B_{RMS}$ as a function of pickup loop diameter. (d) $B_{RMS}$ as function of $h$ for different pickup loop diameters. The effect is larger at low $h$, where $d$ dominates the spatial resolution.

The broadening of the features is clearly visible in Figure 9b and is similar to the broadening due to increasing height. Figure 9c shows how $B_{RMS}$ changes with height for various sizes $d$ of the pickup loop. We can see that, for low values of $h$, the change in $B_{RMS}$ is significant due to $d$ being the dominant factor determining the spatial resolution. For higher values of $h$, the spatial resolution is dominated by $h$, which is visible in the curves merging into a single curve.

Figure 9d shows how $B_{RMS}$ changes as a function of $d$. For higher $d$, the value of $B_{RMS}$ appears to saturate. The limit is of course $B_{RMS} = 0$, if a large enough area is averaged.

## Magnetic Structure Analysis Using Autocorrelation

Another useful tool in analyzing SSM images is 2-dimensional autocorrelation. Calculating the autocorrelation function of an image can provide information about the general structure and distribution of features. In general, the autocorrelation function $R(\delta x, \delta y)$ for some (discrete) signal $B(x, y)$ is given by

$$R(\delta x, \delta y) = \sum_{x,y} B(x + \delta x, y + \delta y)B(x, y). \tag{12}$$

This function will peak if $B$, shifted by $(\delta x, \delta y)$, is similar to the original. Therefore, in the case of periodic structures, calculating the autocorrelation function will give information about the lattice structure. In the case of SSM, this has been used to investigate lattices of superconducting vortices[48].

Additionally, the central peak of the autocorrelation function can be analyzed to get an impression of the dimensions of typical features. The width of a peak along a certain direction will correspond to the size of typical features along that same direction. This way, one can also obtain directional information.

We can apply the 2-dimensional autocorrelation function to a simulated magnetized surface. In Figure 10a, we have simulated a ferromagnetic surface with hexagonal domains that are magnetized with a random orientation in the in-plane direction. When we apply the autocorrelation directly, as shown in Figure 10b, we cannot see any symmetry.

The autocorrelation is a function that identifies periodicity in a signal. Each domain in the simulation creates a positively valued field on one side and a negatively valued field on the other. When these domains are placed next to each other with random orientations (i.e., random magnetic moment direction), the fields can add up or cancel out. Because of the random nature, there will be no periodicity in the field signal, which is why the autocorrelation function returns zero away from the central peak.

We can improve our results by first taking the absolute value of the data in Figure 8a, and then performing the 2-dimensional autocorrelation:

$$R(\delta x, \delta y) = \sum_{x,y} |B(x+\delta x, y+\delta y)||B(x,y)|. \tag{13}$$

This way, the symmetry in the magnetic structure is enhanced by taking positive and negative fields to have the same sign. The results, shown in Figure 10c, clearly show the hexagonal symmetry of the original domain structure. From Figure 10c, we can also find the typical size of the domains, by analyzing the central peak, and the distance between domains, by measuring the peak-to-peak distance. This matches the parameters used for the simulation of Figure 10a.

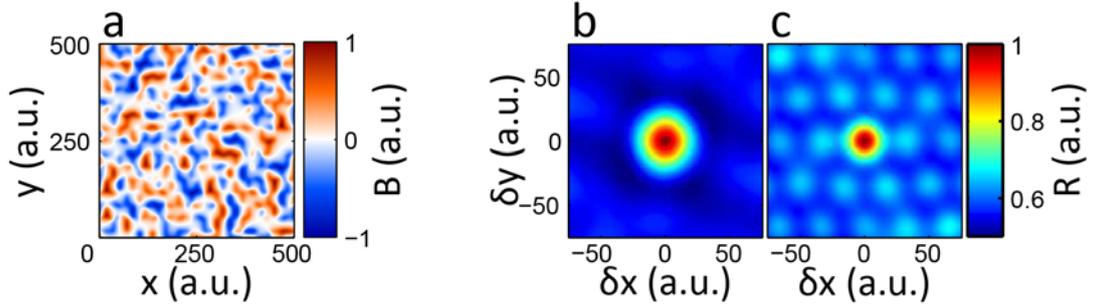

Figure 10. 2D autocorrelation applied to a simulated magnetic surface. (a) Simulated surface of hexagonal domains with magnetic moment aligned randomly in-plane. (b) Center area of the 2-dimensional autocorrelation $R$ of the image in (a). (c) Center area of the 2D autocorrelation of the absolute value of the image in (a). The six-fold symmetry is clearly visible.

A similar analysis can be done using a 2-dimensional Fourier transform, which in the discrete case has the form

$$F(k_x, k_y) = \sum_{x=0}^{\ell_x-1} \sum_{y=0}^{\ell_y-1} e^{-i\left[\left(\frac{2\pi}{\ell_x}\right)k_x x + \left(\frac{2\pi}{\ell_y}\right)k_y y\right]} B(x,y). \tag{14}$$

The Fourier transform and the autocorrelation function are closely related through the Wiener-Khinchin Theorem[49]. A Fourier analysis will give similar results to the autocorrelation, in that it will highlight repetition in the data. On the other hand, it is more limited since it will not give information about the typical dimensions of the domain. As before, the results may be improved by taking the absolute value of $B$ first.

# Outlook
We will now touch on several topics regarding on-going and future developments in SSM.

As mentioned in the introduction, SSM sensors have only recently achieved sub-micrometer dimensions[31, 32]. This trend will continue as fabrication techniques of SQUID sensors improve. This will mean the spatial resolution will improve as well, due to the smaller pickup loop as discussed before. Furthermore, the sensor will also have better field resolution, which can be deduced from the spin sensitivity $S_n$:

$$S_n = \Phi_n \frac{R}{r_e} \left(1 + \frac{h^2}{R^2}\right)^{\frac{3}{2}}, \tag{15}$$

where $\Phi_n$ is the flux noise in $\Phi_0\text{Hz}^{-1/2}$, $R$ is the radius of the pickup loop and $r_e = 2.82 * 10^{-15}$ m is the classic electron radius[32, 50]. Improvements to the spatial and field resolutions will allow for more detailed analysis of certain magnetic phenomena. Examples include single-electron spin measurements, or detailed measurements of Abrikosov vortices or magnetic domain walls. Several fabrication methods have been applied to create nanoSQUIDs, including focused ion-beam milling[32, 51, 52], electron-beam lithography[53] or even carbon-nanotubes[54]. A review on nanoSQUID fabrication and applications was done by Foley and Hilgenkamp[38].

Another topic of interest is local susceptibility measurements. This has been realized by fabricating a secondary coil close to the pickup loop with which a magnetic field can be applied[46]. Such a setup allows for measuring the local susceptibility or otherwise locally manipulating the sample while measuring the resulting magnetic signal.

With the rising interest in topological non-trivial materials, SSM has been used in measuring the edge currents that appear in such systems[44]. The SSM images the magnetic field produced by the edge currents. From this image, the current paths and strength can be calculated.

Some setups have also included the ability to vary the temperature of the sample while imaging[40, 41, 55, 56]. The difficulty here lies with the fact that the SQUID has to be superconducting and therefore has to be kept at cryogenic temperatures. Although measuring at varying temperatures below the sensor's critical temperature is achievable, going to higher temperatures (e.g. room temperature) is more difficult. The solution is to separate the sensor and the sample in space, keeping the sensor in a cryogenic environment and the sample in a system with varying temperature[41, 56]. Because the systems have to be thermally isolated from each other, the distance between the sensor and sample becomes relatively large, meaning the spatial resolution suffers in these systems. The obvious trade-off is the ability to measure magnetism around, for example, the Curie or Néel temperature of samples.

In terms of data analysis, benefits can be found in developing deconvolution methods tailored to the SSM. Since SSM data is a convolution of the actual magnetic flux profile and the pickup loop, the spatial resolution can be improved by applying proper deconvolution. While some early attempts at deconvolution have been made[57, 58], much can be gained by borrowing from other microscopy fields, most notably optical microscopy and astronomy, where deconvolution is actively being used and developed (see, for example, Refs. [59-63]). Accurate deconvolution requires good knowledge of the system geometry, including the sample-sensor separation, which can be difficult to determine with high precision.

## Conclusion

In this report we have discussed imaging magnetism using SSM. We have covered several aspects that contribute to how SSM images are formed and what information can be obtained from them. Here we will summarize the most important results.

We started by looking at the spatial resolution of an SSM system. We observed that both the scanning height and the pickup loop diameter influence the resolution. Therefore, either parameter on its own is not sufficient to properly describe the spatial resolution of an SSM setup. We have defined the spatial resolution to be the separation between the two extrema of an imaged in-plane point-dipole. We believe this definition is intuitive and simple, and incorporates both height and the dimensions of the pickup loop.

Next, we discussed imaging magnetic features using SSM. Using simulations, we have seen how dipoles will be imaged in different orientations and if multiple dipoles are close together. From there, we looked at imaging ferromagnetic surfaces. We noticed how the parameter $B_{RMS}$ sharply drops with increasing height, resulting in much of the information about the underlying domain structure being lost. We also looked at how the strength and orientation of the magnetic moment influences the final image. The difference between in-plane and out-of-plane magnetic moment is only visible at low scanning height. Beyond that, we observed a linear relation between $B_{RMS}$ and the magnitude of the magnetic moment, with a larger coefficient for the out-of-plane case.

Finally, we have shortly discussed using 2-dimensional autocorrelation to extract more information about the structure of the imaged magnetic phenomena. Using a simulated array of ferromagnetic hexagons, we showed that this function can be used to find typical feature size, orientation and distance to neighboring features.

## Acknowledgements

This work is part of the research program of the Netherlands Organisation for Scientific Research (NWO). X.R.W. thanks support of NWO Rubicon grants (2011, 680-50-1114), Elite Nanyang Assistant Professorship grant from Nanyang Technological University and Academic Research Fund Tier 1 from Singapore Ministry of Education.